\documentclass[aps,prb,amsmath,amssymb,footinbib,showpacs,twocolumn,superscriptaddress,longbibliography]{revtex4-2}

\usepackage{graphicx}
\usepackage{braket}
\usepackage{mathrsfs}
\usepackage{hyperref}
\usepackage{balance}
\usepackage{color}
\usepackage{float}
\usepackage[caption=false]{subfig}

\newcommand{\lrp}[1]{\left(#1\right)}

\begin{document}

\title{Pfaffian-based topological invariants for one dimensional semiconductor-superconductor heterostructures}

\author{Binayyak B. Roy}
\affiliation{Department of Physics and Astronomy, Clemson University, Clemson, SC 29634, USA}

\author{William B. Cason}
\affiliation{Department of Physics and Astronomy, Clemson University, Clemson, SC 29634, USA}

\author{Nimish Sharma}
\affiliation{Department of Physics, BITS Pilani-Pilani Campus, Rajasthan, 333031, India}

\author{Sumanta Tewari}
\affiliation{Department of Physics and Astronomy, Clemson University, Clemson, SC 29634, USA}


\begin{abstract}
We review the Pfaffian-based $\mathbb{Z}_2$ topological invariants in one dimensional semiconductor–superconductor (SM-SC) nanowire heterostructures and clarify their validity in finite and disordered systems. For the clean nanowire, the product of the Pfaffians of the Hamiltonian at particle–hole symmetric momenta $k=0,\pi$ changes sign at the topological phase transition defined by the bulk gap closing, leading to the definition of $\mathbb{Z}_2$ Kitaev invariant also known as Majorana number. We show that this momentum-space invariant is equivalent to a real space construction based on twisted boundary conditions, in which the sign of the product of the Pfaffians of the Hamiltonian under periodic and anti-periodic boundary conditions defines the $\mathbb{Z}_2$ index. By introducing a superlattice description of periodically repeated disorder, we demonstrate that the real space Pfaffian invariant defined as the sign of the Pfaffians of the Hamiltonian with periodic and anti-periodic boundary conditions, remains a well defined invariant even in the absence of microscopic translational symmetry. Within this framework, it is also equivalent to the recently defined periodic disorder invariant (PDI), which constitutes an integer valued ($\mathbb{Z}$) topological invariant in the presence of chiral symmetry. Finally, we prove that the sign of the Pfaffian of a quadratic Hamiltonian equals the fermion parity of its ground state, establishing a direct physical interpretation of the invariant, in terms of sign of the product of the ground state fermion parity with periodic and anti-periodic boundary conditions. Numerical results confirm the correspondence between sign of the Pfaffian reversals, flux-induced level crossings, and ground-state parity switching in clean and disordered nanowires.
\end{abstract}

\maketitle

\section{Introduction}

One dimensional SM-SC hybrid nanowires with spin–orbit coupling and Zeeman splitting provide a well established platform for realizing topological superconductivity and Majorana zero modes\cite{sau2010non,sau2010generic,alicea2010majorana,tewari2010theorem,lutchyn2010majorana,oreg2010helical,lutchyn2011search,alicea2011non,hassler2010anyonic,tewari2011topologically,mao2011superconducting,mao2012hole,stanescu2011majorana,qu2011signature,lutchyn2012momentum,van2012coulomb}. These systems offer a concrete setting in which topological concepts can be directly related to experimentally accessible observables. Central to this connection is the notion of a topological invariant, a discrete quantity that remains unchanged under smooth deformations of the Hamiltonian as long as the bulk excitation gap remains finite, and encodes robust information about ground state and boundary properties.

Among the invariants used to characterize one dimensional topological superconductors \cite{PhysRevB.83.155429,tewari2012topologicaalinvariant}, the Pfaffian invariant introduced by Kitaev \cite{Kitaev2001} occupies a central role. Defined for particle–hole symmetric Bogoliubov–de Gennes (BdG) Hamiltonians, the Pfaffian invariant defined as the sign of the product of the Pfaffians of the Hamiltonian at $k=0$ and $k=\pi$ provides a $\mathbb{Z}_2$ classification. This distinguishes topologically trivial and nontrivial superconducting phases and is directly related to the presence or absence of Majorana zero modes at the ends of an open wire. 
A key insight underlying the Pfaffian invariant is its intimate connection to the fermion parity of the superconducting ground state\cite{Kitaev2001}. In one dimensional systems, changes in the sign of the Pfaffian of the Hamiltonian are accompanied by switches of the ground state fermion parity and, necessarily, coincide with bulk gap closings or protected low energy level crossings. From this perspective, the Pfaffian invariant is not merely an abstract topological quantity, but a diagnostic of physically meaningful parity changing processes that may be probed through flux dependent measurements \cite{PRBSau2025,ArXivStanescu2025,MSRMicrosoft2024}.

Several real space approaches have been developed to define topological invariants for finite and spatially inhomogeneous SM-SC nanowire systems \cite{PhysRevB.83.155429,eissele2025topological,PhysRevB.110.115436,FulgaPRB2012,PRBSau2025,DasPRB2016, DasPRB2023, DasPRBSpectra2023}, where microscopic translational symmetry is absent. A particularly transparent formulation is based on boundary conditions rather than crystal momentum. In this formulation the Kitaev invariant, defined as the sign of the product of the Pfaffian of the Hamiltonian at $k=0, \pi$, can be shown to be equivalent to the sign of the product of the Pfaffians of the Hamiltonian with periodic and anti-periodic boundary conditions. In a finite superconducting wire, imposing periodic or anti-periodic boundary conditions is equivalent to threading a magnetic flux through a ring formed by joining the wire’s ends\cite{PhysRevB.99.035312, LutchynPRL2010}. The flux-induced phase twist modifies the boundary hopping and thereby provides a physically transparent means of continuously deforming the Hamiltonian. Within this framework, the product of Pfaffians evaluated under periodic and anti-periodic boundary conditions plays a role analogous to evaluating the Pfaffian at particle–hole symmetric momenta in the clean, translationally invariant limit, while remaining well defined in real space.

An alternative real space construction in the presence of disorder is provided by the superlattice formulation\cite{eissele2025topological}, in which a finite wire is embedded into an enlarged periodic structure by repeating in a superlattice that captures the full spatial profile of the system. This procedure restores translational symmetry at the level of the superlattice and enables the definition of Pfaffian-based invariants in terms of superlattice Bloch momenta, even in the absence of microscopic translational invariance. Following this approach, the recently discussed periodic disorder invariant (PDI)\cite{eissele2025topological}, has proven useful for diagnosing topology in spatially inhomogeneous systems. It can be shown that the Pfaffian invariant defined in the space of superlattice Bloch momentum, as the sign of the product of the Pfaffian of the superlattice Hamiltonian at particle-hole symmetric superlattice Bloch momenta $q=0, \pi$, is equivalent to the sign of the product of the Pfaffian of the Hamiltonian of the finite wire with periodic and anti-periodic boundary conditions. Thus, from the superlattice formulation of the topological invariant in the disordered systems it follows that the sign of the product of the Pfaffian of the Hamiltonian of a finite disordered SM-SC nanowire with periodic and anti-periodic boundary conditions remains as a well defined topological invariant. To avoid ambiguity when referring to different formulations of the Pfaffian invariant, we will distinguish between three closely related Pfaffian constructions used throughout this work. The momentum-space Pfaffian invariant refers to the sign of the product of Pfaffians evaluated at the particle–hole symmetric momenta $k=0$ and $k=\pi$. The twisted-boundary Pfaffian invariant denotes the corresponding real-space quantity defined from the Pfaffians of the Hamiltonian under periodic and anti-periodic boundary conditions. Finally, the superlattice Pfaffian invariant refers to the Pfaffian construction evaluated at the particle–hole symmetric superlattice Bloch momenta $q=0$ and $q=\pi$. As we show below, these seemingly distinct formulations are equivalent and provide complementary perspectives on the same $\mathbb{Z}_2$ topological classification.

The purpose of this work is to present a unified and physically transparent framework for Pfaffian-based topological invariants in one dimensional SM-SC hybrid systems. Starting from Kitaev’s original momentum-space construction, we establish its equivalence to a real space formulation based on twisted boundary conditions. This strongly indicates that even in the presence of disorder, the twisted-boundary Pfaffian may provide a well defined $\mathbb{Z}_2$ invariant. We then extend the analysis to spatially inhomogeneous systems through a superlattice construction. Within this framework, we show that the superlattice Pfaffian invariant is equivalent to twisted-boundary Pfaffian invariant of a finite SM-SC nanowire. We further show that the superlattice Pfaffian invariant is directly related to the parity of the PDI. Finally, we establish the general relation between the Pfaffian of a quadratic Hamiltonian and the fermion parity of its ground state, and show how changes in the sign of the Pfaffian track flux-induced level crossings in finite and disordered nanowire systems. By placing these results within a single coherent framework, we aim to clarify the conceptual foundations of Pfaffian invariants and provide practical examples for their application in experimentally relevant topological superconducting devices.

The remainder of this paper is organized as follows. In Sec.~\ref{sec:momentum_clean}, we review the Pfaffian formulation~\cite{Kitaev2001} of the $\mathbb{Z}_2$ invariant in momentum space and clarify why the topological classification is completely determined by particle–hole symmetric momenta. We then apply this framework to the clean SM-SC nanowire model and derive the corresponding topological phase transition criterion. In Sec.~\ref{sec:relation_bw_ms_rs}, we reformulate the invariant in real space by introducing twisted boundary conditions within the tight-binding representation and establish its equivalence to the momentum-space construction. Sec.\ref{sec:PDI} extends this analysis to spatially inhomogeneous systems through the superlattice formulation and the associated periodic disorder invariant. Sec.~\ref{sec:proof_fermion_parity}, presents a general proof that relates the Pfaffian of a quadratic Hamiltonian in the Majorana basis to the fermion parity of its ground state, thereby providing a direct physical interpretation of the invariant. Finally, in Sec.~\ref{sec:numerical_results}, we present numerical results that illustrate the equivalence between Pfaffian-based invariants, their real space and superlattice formulations, and flux-induced fermion parity switching in finite and disordered systems.

\section{Momentum-space Pfaffian invariant and topological criterion in a clean nanowire}
\label{sec:momentum_clean}

We begin by reviewing the Pfaffian formulation of the $\mathbb Z_2$ topological invariant for a translationally invariant quadratic Hamiltonian. Consider a one dimensional BdG Hamiltonian written in momentum space,
\begin{equation}
H = \frac{1}{2}\sum_k \Psi_k^\dagger \,\mathcal H(k)\,\Psi_k,
\label{eq:moment_Hamilt}
\end{equation}
defined in a Nambu basis in which $\mathcal H(k)$ satisfies particle–hole symmetry,
\begin{equation}
\mathcal C\,\mathcal H(k)\,\mathcal C^{-1}
=
-\mathcal H(-k),
\qquad
\mathcal C=\tau_x K.
\end{equation}

\noindent Particle–hole symmetry enforces spectral symmetry about zero energy, so that eigenvalues occur in $\pm E$ pairs.

\noindent Using particle–hole symmetry, we construct the matrix
\begin{equation}
B(k) = \mathcal H(k)\,\tau_x \qquad B(k)^T = - B(-k)
\end{equation}
Consequently, at momenta satisfying
\begin{equation}
k=-k \quad (\mathrm{mod}\ 2\pi),
\end{equation}
the matrix $B(k)$ becomes antisymmetric,
\begin{equation}
B(k)^T=-B(k).
\end{equation}
In one dimension, these particle–hole symmetric momenta are
\begin{equation}
k=0,\quad k=\pi
\end{equation}
for lattice systems defined in a reduced Brillouin zone.

At these special points, the antisymmetry of $B(k)$ ensures that its Pfaffian is well defined. For generic momenta with $k\neq -k$, the points $k$ and $-k$ form distinct sectors related by particle–hole symmetry. When the Hamiltonian is organized in the combined $(k,-k)$ subspace, the corresponding blocks contribute to the Pfaffian of the full antisymmetric matrix through positive factors proportional to $\det B(k) \sim \prod_j E_j(k)^2$. Since these contributions are strictly nonnegative, they cannot alter the overall sign of the Pfaffian. By contrast, at $k=0$ and $k=\pi$, each momentum point maps onto itself under particle–hole symmetry. At these momenta, an odd number of eigenvalue crossings through zero changes the sign of the Pfaffian. The $\mathbb Z_2$ topology is therefore completely determined by the particle–hole symmetric points.

\noindent The topological invariant may thus be written as \cite[Eq. 26]{Kitaev2001}
\begin{equation}
\mathrm{Pf_{inv}} =\mathrm{sgn}\left[\mathrm{Pf}(B(0))\mathrm{Pf}(B(\pi))\right]
\label{eq:Z2_pfaffian}
\end{equation}

A change in $\mathrm{Pf_{inv}}$ requires that one of the Pfaffians vanish, which occurs precisely when the bulk gap closes at $k=0$ or $k=\pi$. The Pfaffian invariant is therefore fixed by the behavior of the Hamiltonian at the particle–hole symmetric momenta.

\vspace{0.3cm}

We now specialize this general construction to the one dimensional semiconductor nanowire with Rashba spin–orbit coupling, proximity-induced $s$-wave pairing, and a Zeeman field. In momentum space, the system is described by the BdG Hamiltonian in Eq.~\ref{eq:moment_Hamilt} written in the standard Nambu basis
\begin{equation}
\Psi_k =
\left(
c_{k\uparrow},
c_{k\downarrow},
c_{-k\uparrow}^\dagger,
c_{-k\downarrow}^\dagger
\right)^T.
\end{equation}

\noindent In the low-$k$ limit, the BdG kernel takes the compact form \cite{tewari2010theorem}
\begin{equation}
\mathcal{H}(k)
=
\xi_k \tau_z \sigma_0
+
\alpha k \, \tau_z \sigma_y
+
\Gamma \, \tau_z \sigma_x
-
\Delta \, \tau_y \sigma_y,
\label{eq:BdG_standard}
\end{equation}
where $\xi_k = \frac{k^2}{2m} - \mu$, $\mu$ is the chemical potential, $\alpha$ denotes the Rashba coupling, $\Gamma$ is the Zeeman field, and $\Delta$ is the induced superconducting pairing amplitude. The Pauli matrices $\sigma_i$ act in spin space and $\tau_i$ act in particle–hole space. All parameters are taken to be real. The BdG Hamiltonian satisfies particle–hole symmetry,
\begin{equation}
\mathcal{C} \, \mathcal{H}(k) \, \mathcal{C}^{-1}
=
-
\mathcal{H}(-k),
\qquad
\mathcal{C} = \tau_x \sigma_0 K,
\end{equation}
with $\mathcal{C}^2 = +1$, placing the system in symmetry class D \cite{tewari2012topologicaalinvariant}. The spectrum is therefore symmetric about zero energy.

The topological phase transition is determined by a bulk gap closing. Because the spin–orbit term vanishes at $k=0$, the gap closing generically occurs at $k=0$. Evaluating Eq.~\eqref{eq:BdG_standard} at $k=0$ yields
\begin{equation}
\mathcal{H}(0)
=
-\mu \tau_z \sigma_0
+
\Gamma \tau_z \sigma_x
-
\Delta \tau_y \sigma_y.
\end{equation}

\noindent The eigenvalues at $k=0$ are
\begin{equation}
E(0)
=
\pm \left| \Gamma \pm \sqrt{\mu^2 + \Delta^2} \right|.
\end{equation}
The smallest positive eigenvalue vanishes when
\begin{equation}
\Gamma^2 = \mu^2 + \Delta^2.
\end{equation}
Thus, the bulk gap closes at
\begin{equation}
\boxed{
|\Gamma_c| = \sqrt{\mu^2 + \Delta^2}.
}
\label{eq:critical_field}
\end{equation}

For $|\Gamma| > \sqrt{\mu^2 + \Delta^2}$ the system is in the topological phase supporting Majorana zero modes at its ends, whereas for $|\Gamma| < \sqrt{\mu^2 + \Delta^2}$ it is topologically trivial.

In the continuum limit, the $k=\pi$ contribution to Eq.~\eqref{eq:Z2_pfaffian} is trivial \cite{oreg2010helical,tewari2012topologicalminigap,stanescu2011majorana}, and the invariant reduces to the sign of the Pfaffian at $k=0$. Evaluating the Pfaffian yields
\begin{equation}
\mathrm{Pf}(B(0))=
\Gamma^2 - (\mu^2 + \Delta^2)
\label{eq:pfaf_tqpt}
\end{equation}
Hence, the invariant changes sign precisely at the bulk gap-closing condition~\eqref{eq:critical_field}.

\noindent It is instructive to contrast the Pfaffian criterion with other seemingly natural spectral diagnostics. In particular, one might ask whether $\det \mathcal{H}(k)$ could itself be used to define a topological invariant. Particle–hole symmetry ensures that
\begin{equation}
\det \mathcal{H}(k)
=
\prod_n E_n(k)^2 \ge 0,
\end{equation}
which is manifestly nonnegative and therefore contains no sign information distinguishing distinct topological sectors.

Moreover, writing $B(k)=\mathcal H(k)\,\tau_x$, one finds
\begin{equation}
\det \mathcal{H}(k)
=
\left[\prod_{k=-k} \mathrm{Pf}(B(k))\prod_{k\neq-k}\det B(k)\right]^2
\end{equation}
While $\det B(k)$ is always nonnegative, the factor $\displaystyle \prod_{k=-k} \mathrm{Pf}(B(k))$ enters only through its square. The determinant therefore eliminates the sign structure encoded in the Pfaffian. It is precisely this sign that changes when an odd number of eigenvalues crosses zero (refer to Eq.~\ref{eq:pfaf_tqpt}).

Accordingly, the Pfaffian evaluated at particle–hole symmetric momenta provides a natural $\mathbb{Z}_2$ invariant directly tied to bulk gap closings. Having established the momentum-space criterion for the topological transition, we now reformulate this invariant in real space by relating Pfaffians at particle–hole symmetric momenta to twisted boundary conditions in the tight-binding nanowire model.

The Pfaffian evaluated at particle–hole symmetric momenta $k=0,\pi$, therefore provides a natural $\mathbb{Z}_2$ topological invariant for translationally invariant 1D SM-SC nanowire systems. In the clean SM–SC nanowire, the changes in the sign of the Pfaffian occur precisely at the bulk gap closing condition $|\Gamma_c|=\sqrt{\mu^2+\Delta^2}$, establishing the correspondence between the Pfaffian invariant and the topological phase transition of the system.

\section{Real space formulation of the Pfaffian invariant}
\label{sec:relation_bw_ms_rs}

In the previous section we showed that the topology of a translationally invariant nanowire is determined by the sign of the product of Pfaffians evaluated at the particle–hole symmetric momenta $k=0$ and $k=\pi$. We now demonstrate that an equivalent invariant may be defined in real space by imposing twisted boundary conditions on a finite nanowire. Specifically, we show that the sign of the product of Pfaffians of the Hamiltonian evaluated under periodic and anti-periodic boundary conditions is equivalent to the momentum-space Pfaffian invariant.

To formulate the Pfaffian invariant directly in real space, we consider the same one dimensional SM-SC nanowire with its ends joined to form a topological superconducting ring threaded by a magnetic flux $\Phi$\cite{PhysRevLett.7.46,PhysRevLett.89.096802,PhysRevB.109.064504}. In this geometry, the Hamiltonian takes the form

\balance
\begin{widetext}

\begin{equation}
    \mathcal{H} = \frac{1}{2}\sum_n\Psi^\dagger_n\mathrm{H_{onsite}}\Psi_n + \frac{1}{2}\sum_{i,j}\Psi^\dagger_i\mathrm{H_{hopping}}\Psi_j\delta_{(i,j),(i,i+1)} + \left(\Psi^\dagger_0\mathrm{H_{ends}}(\phi)\Psi_{L-1} + h.c.\right)
    \label{eq:single_chain_H}
\end{equation}

\end{widetext}

\noindent where $\Psi^\dagger_n=(c^\dagger_{n\uparrow},c^\dagger_{n\downarrow},c_{n\uparrow},c_{n\downarrow})$ denotes the real space Nambu spinor. The onsite term is

\begin{equation}
    \mathrm{H_{onsite}} = (2t-\mu)\tau_z\sigma_0 + \Gamma\tau_z\sigma_x - \Delta\tau_y\sigma_y,
\end{equation}

\noindent where $t$ is the nearest-neighbor hopping amplitude, while $\mu$, $\Gamma$, and $\Delta$ retain their meanings from the continuum formulation. The nearest-neighbor hopping term is

\begin{equation}
    \mathrm{H_{hopping}} = -t\tau_z\sigma_0 + \frac{i}{2}\alpha\tau_z\sigma_y,
    \label{eq:hop}
\end{equation}

\noindent with $\alpha$ the Rashba spin–orbit coupling strength. The boundary term,

\begin{equation}
    \mathrm{H_{ends}}(\phi) = \lambda e^{i\phi}\left[-t\tau_z\sigma_0 + \frac{i}{2}\alpha\tau_z\sigma_y\right],
\end{equation}

\noindent describes the flux-dependent coupling between the ends of the nanowire, where $\lambda$ controls the strength of the boundary link. A magnetic flux $\Phi$ threading the ring is incorporated through a Peierls phase on this boundary link, which enforces a boundary twist $\phi=\pi\Phi/\Phi_0$ ($\Phi_0=h/2e$).

A gauge transformation shifts the flux-induced phase $\phi=\pi\Phi/\Phi_0$, with $\Phi_0=h/2e$, from the boundary to the bulk,

\begin{equation}
    U_\phi : \tilde{c}_{j,\sigma} \mapsto e^{i\phi j/L}c_{j,\sigma},
\end{equation}

\noindent where $j\in[0,L-1]$ labels lattice sites. In this gauge, and in the absence of boundaries, translational invariance is restored, $\mathcal{H}_{i,j}=\mathcal{H}_{i+1,j+1}$, allowing block diagonalization in momentum-space. The Fourier transformation reads

\begin{equation}
    \tilde{c}_{j,\sigma} = \frac{1}{\sqrt{L}}\sum_{k_n}e^{ik_n(\phi)j}c_{k_n(\phi),\sigma},
\end{equation}

\noindent where the quantized momenta are $k_n(\phi)=(2\pi n+\phi)/L$.

For an infinite, translationally invariant system, the Pfaffian invariant is defined as

\begin{equation}
    \mathrm{Pf_{inv}} = \mathrm{sgn}[\mathrm{Pf({\mathcal{H}(k=0)})}\mathrm{Pf(\mathrm{\mathcal{H}(k=\pi)})}].
\end{equation}

A bulk gap closing can occur only at the particle–hole symmetric points $k=0$ and $k=\pi$ as shown in Sec.~\ref{sec:momentum_clean}. In the presence of a flux-induced phase distributed throughout the bulk, the allowed momenta become

\begin{equation}
    k_n(\phi) = \frac{2\pi n+\phi}{L},
\end{equation}

\noindent with $n\in(0,2,\ldots,L-1)$. The relation between boundary conditions and particle–hole symmetric momenta depends on whether $L$ is even or odd.

\begin{itemize}
    \item[I.] For even $L$, $k_n(\phi=0)=2\pi n/L$ (periodic boundary conditions) includes both particle–hole symmetric points $\{0,\pi\}$ for $n=(0,L/2)$. In contrast, $k_n(\phi=\pi)=2\pi(n+1/2)/L$ (anti-periodic boundary conditions) excludes both. This implies
    \begin{align}
        & \mathrm{sgn}[\mathrm{Pf({\mathcal{H}(\phi=0)})}] \equiv \nonumber \\
        & \mathrm{sgn}[\mathrm{Pf({\mathcal{H}(k=0)})}\mathrm{Pf(\mathrm{\mathcal{H}(k=\pi)})}].
    \end{align}

    \item[II.] For odd $L$, $k_n(\phi=0)=2\pi n/L$ includes only $k=0$, while $k_n(\phi=\pi)=2\pi(n+1/2)/L$ includes only $k=\pi$. Accordingly,
    \begin{align}
        & \mathrm{sgn}[\mathrm{Pf({\mathcal{H}(\phi=0)})}] \equiv \mathrm{sgn}[\mathrm{Pf({\mathcal{H}(k=0)})}], \nonumber \\
        & \mathrm{sgn}[\mathrm{Pf(\mathrm{\mathcal{H}(\phi=\pi)})}] \equiv \mathrm{sgn}[\mathrm{Pf(\mathrm{\mathcal{H}(k=\pi)})}].
    \end{align}
\end{itemize}

Therefore, in the clean system, the boundary twists $\phi=0$ and $\phi=\pi$ isolate the particle–hole symmetric sectors corresponding to $k=0$ and $k=\pi$, with the precise identification determined by the parity of $L$. The product of Pfaffians evaluated under periodic and anti-periodic boundary conditions thus provides a real space representation of the $\mathbb{Z}_2$ invariant. The two cases may be combined into

\begin{align}
        & \mathrm{sgn}[\mathrm{Pf({\mathcal{H}(\phi=0)})}\mathrm{Pf(\mathrm{\mathcal{H}(\phi=\pi)})}] \equiv \nonumber \\
        & \mathrm{sgn}[\mathrm{Pf({\mathcal{H}(k=0)})}\mathrm{Pf(\mathrm{\mathcal{H}(k=\pi)})}].
\end{align}

This boundary condition formulation yields a real space characterization of the Pfaffian invariant that remains well defined in finite systems. We have therefore shown that the product of Pfaffians of the Hamiltonian evaluated under periodic ($\phi=0$) and anti-periodic ($\phi=\pi$) boundary conditions is equivalent to the product of Pfaffians evaluated at the particle–hole symmetric momenta $k=0$ and $k=\pi$ and thus acts as a real space $\mathbb{Z}_2$ invariant for 1D SM-SC nanowire systems. This establishes a real-space representation of the Pfaffian invariant for finite SM–SC nanowires.

\begin{equation}
    \mathrm{Pf}_{\mathrm{inv}}^\phi = \mathrm{sgn}[\mathrm{Pf({\mathcal{H}(\phi=0)})}\mathrm{Pf(\mathrm{\mathcal{H}(\phi=\pi)})}]
    \label{eq:real_sp_Pfaff}
\end{equation}

\section{Superlattice Formulation and Periodic Disorder Invariant}
\label{sec:PDI}

In the previous section we established that the momentum-space Pfaffian invariant can be expressed in real space through twisted boundary conditions, where the sign of the product of Pfaffians evaluated under periodic and anti-periodic boundary conditions reproduces the topological invariant defined at the particle-hole symmetric momenta $k=0$ and $k=\pi$. However, this derivation relies implicitly on translational invariance. In spatially inhomogeneous systems, where disorder breaks microscopic translational symmetry and crystal momentum is no longer a good quantum number, the applicability of momentum-space formulation is no longer viable. By contrast, the real space formulation based on twisted boundary conditions can still be defined even in the presence of disorder. However, once translational symmetry is broken, it is not immediately evident that the twisted-boundary Pfaffian invariant construction continues to define a bona fide topological invariant.

To address this issue, we introduce a superlattice construction in which the disorder profile of a finite nanowire is periodically repeated, thereby restoring translational symmetry at the level of the supercell. Within this framework, the Bogoliubov--de Gennes Hamiltonian may be expressed in terms of a superlattice Bloch momentum $q$, allowing the Pfaffian invariant to be formulated in direct analogy with Kitaev's treatment of closed chains \cite{Kitaev2001} in defining the Majorana number. We then compare this superlattice Pfaffian invariant with the periodic disorder invariant (PDI), which arises when an additional chiral symmetry is present \cite{tewari2012topologicaalinvariant} and has previously been shown to provide a useful diagnostic of topology in spatially inhomogeneous systems \cite{eissele2025topological}. We then show that, in the special case of a single supercell, the superlattice Bloch formulation reduces precisely to the real space twisted boundary condition construction, even in the presence of disorder. This establishes that the real space Pfaffian invariant defined through periodic and anti-periodic boundary conditions remains a valid $\mathbb{Z}_2$ invariant for the disordered one dimensional SM-SC hybrid system.

\begin{figure}[h]
    \centering
    \includegraphics[width=\linewidth]{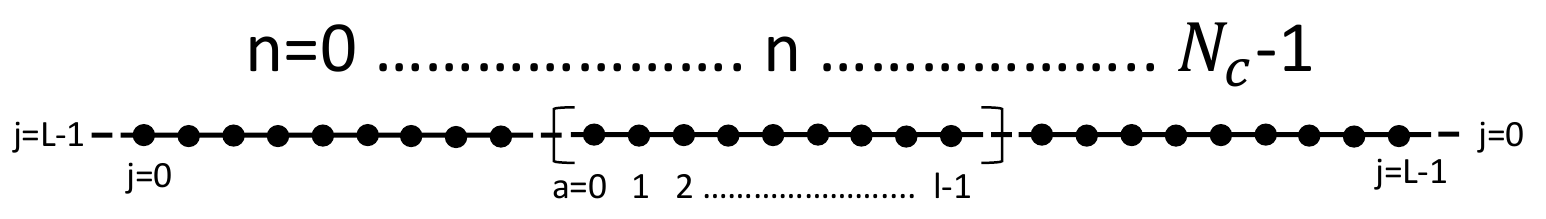}
    \caption{Schematic illustration of the superlattice formulation and indexing convention used throughout the manuscript. The global site index $j\in[0,L-1]$ is decomposed into a supercell index $n\in[0,N_c-1]$ and a local index $a\in[0,l-1]$, with $j = nl + a$, where $l$ denotes the number of microscopic sites per supercell.}
    \label{fig:Sup_latt}
\end{figure}

We introduce a superlattice description in which each supercell contains the full disorder profile within its extent. Translational symmetry is then restored on the scale of the supercell, such that $V_{j+l} = V_j$, where $l$ denotes the number of microscopic sites within a single supercell. The global lattice index $j$ is written as $j = nl + a$, with $n$ labeling the supercell and $a$ denoting the local index of a microscopic site within the supercell (as shown in Fig.~\ref{fig:Sup_latt}).

\noindent Within this superlattice description, the Hamiltonian and the corresponding BdG basis must be reformulated in terms of supercell degrees of freedom.

At a microscopic site $j$, the BdG basis is given by $\psi^\dagger_j = (c^\dagger_{j\uparrow},c^\dagger_{j\downarrow},c_{j\uparrow},c_{j\downarrow})$. For a given supercell labeled by $n$, the corresponding supercell spinor is constructed by stacking the $l$ microscopic sites within the supercell,
\[
\Psi^\dagger_n = (\psi^\dagger_{n,0}, \psi^\dagger_{n,1}, \psi^\dagger_{n,2}, \ldots, \psi^\dagger_{n,l-1}),
\]
so that $\Psi^\dagger_n$ has dimension $4l$. Consequently, the superlattice Bloch Hamiltonian associated with each supercell is a $4l \times 4l$ matrix. The full real space Hamiltonian of the superlattice then takes the form

\balance
\begin{widetext}
    \begin{equation}
    H = \frac{1}{2}\sum_n\left[\Psi^\dagger_n\mathrm{H_0}\Psi_n + \Psi^\dagger_{n+1}\mathrm{H_1}\Psi_n+\Psi^\dagger_{n}\mathrm{H^\dagger_1}\Psi_{n+1}\right]
    \label{eq:suplatt_no_phase}
\end{equation}
\end{widetext}

\noindent where $\mathrm{H_0}$ contains the onsite terms and the intra-supercell hopping contributions between neighboring microscopic sites $j$ and $j+1$, while $\mathrm{H_1}$ describes the inter-supercell hopping that couples supercell $n$'s last site to supercell $n+1$'s first site (using the same $\mathrm{H_{hopping}}$ matrix as in Eq.~\ref{eq:hop}). For a superlattice consisting of $N_c$ supercells, we impose periodic boundary conditions by identifying the first and last supercells of the lattice. 
\begin{equation}
    \Psi_{n+N_c} = \Psi_n
\end{equation}
This allows the system to be treated as translationally invariant under discrete shifts by one supercell.

\noindent The superlattice construction restores translational invariance under discrete shifts by one supercell of length $l$ even in the presence of disorder. Consequently, the Hamiltonian is invariant under the supercell translation operator $T_l$.

Under this symmetry, Bloch's theorem applies with respect to supercell translations, and eigenstates may be labeled by a conserved superlattice Bloch momentum $q$. The Hamiltonian may therefore be expressed in the corresponding Bloch representation. In direct analogy with the momentum-space formulation discussed in Sec.~\ref{sec:momentum_clean}, the topology of the system can then be characterized by evaluating the Pfaffian of the Hamiltonian at the particle--hole symmetric points of the superlattice Brillouin zone.

The resulting conserved quantum number is the superlattice Bloch momentum $q$,
\begin{equation}
    q = \frac{2\pi n}{N_c}, \qquad n=0,1,\ldots,N_c-1.
    \label{eq:bloch_moment}
\end{equation}

\noindent This construction allows the BdG Hamiltonian to be expressed in terms of the superlattice Bloch momentum $q$ by Fourier transforming the supercell operators,

\begin{equation}
\Psi_{q} = \frac{1}{\sqrt{N_c}} \sum_n e^{iqn}\Psi_n,
\end{equation}

\noindent where the allowed Bloch momenta are $q$ as shown in Eq.~\ref{eq:bloch_moment}. In this representation the BdG Hamiltonian decomposes into independent blocks labeled by ${q}$ \cite{eissele2025topological}.

\begin{eqnarray}
H &=& \frac{1}{2}\sum_{q} \Psi_q^\dagger\,\widehat{H}(q)\,\Psi_q \nonumber\\
\widehat{H}_{ij}(q) &=& [h_0 + V_{dis}(i)~\!\sigma_0\tau_z]~\! \delta_{ij}+h_1~\! \delta_{j,i+1}+h_1^\dagger~\! \delta_{j,i-1} \nonumber \\
&+& h_1e^{i q}~\! \delta_{i,l}\delta_{j,1}+h_1^\dagger e^{-i q}~\! \delta_{i,1}\delta_{j,l}, \label{hatH}
\end{eqnarray}

\noindent where $h_0$ is equivalent to $H_{\mathrm{onsite}}$ and $h_1$ is equivalent to $H_{\mathrm{hopping}}$. Following the argument used for translationally invariant systems, the topology of the superlattice Hamiltonian is determined by the Pfaffian of the Hamiltonian evaluated at the particle–hole symmetric points of the superlattice reduced Brillouin zone. The corresponding superlattice Pfaffian invariant is therefore determined by the sign of the product of Pfaffians of the Hamiltonian evaluated at $q=0$ and $q=\pi$.

\begin{equation}
    \mathrm{Pf_{inv}^q} = \mathrm{sgn}\left[\mathrm{Pf}(\widehat{H}(q=0))\mathrm{Pf}(\widehat{H}(q=\pi))\right]
    \label{eq:sup_latt_Pf}
\end{equation}

When the Hamiltonian possesses an additional chiral symmetry, an alternative topological invariant known as the periodic disorder invariant (PDI), a $\mathbb{Z}$ invariant, may be defined for spatially inhomogeneous systems. The PDI has previously been shown to provide a reliable diagnostic of topology in disordered nanowire systems \cite{eissele2025topological}. It is therefore instructive to compare this invariant with the Pfaffian invariant formulated in the superlattice Bloch momentum space. The periodic disorder invariant for the Hamiltonian defined in Eq.~\ref{hatH} is

\begin{equation}
\nu = \frac{1}{2}{\rm Tr}\left[\mathcal{S}(h_1\overline{g}_1^\dagger-h_1^\dagger \overline{g}_1)\right], \label{nu}
\end{equation}

\noindent where $h_1$ denotes the nearest-neighbor (intra-wire) hopping matrix, and $\overline{g}_1$ is the position-averaged nearest-neighbor Green's function, defined and discussed in detail in Ref.~\cite{eissele2025topological}. Since all block terms remain proportional to $\tau_z$ and $\tau_y$, the Hamiltonian $\widehat{H}(q)$ anticommutes with $\tau_x$, thereby preserving chiral symmetry ($\mathcal{S}=\tau_x\sigma_0$). Owing to this chiral symmetry, the Hamiltonian may be off-diagonalized in a manner analogous to the clean-system case (using $\mathcal{S}$ or, equivalently, the unitary rotation $U=e^{i\frac{\pi}{4}\tau_y}$), with the distinction that the construction is now carried out in the superlattice Bloch momentum $q$-space and is relevant only at the particle-hole symmetric points $q=0,\pi$,
\begin{equation}
    U\widehat{H}(q)U^\dagger = \begin{pmatrix}
        0 & A(q) \\
        A^T(-q) & 0
    \end{pmatrix}.
\end{equation}

\noindent Here $A(q)$ denotes the momentum-space representation of $A$, the off-diagonal block of $\widehat{H}(q)$ written in skew-symmetric form. The determinant $\mathrm{Det}(A(q))$ can vanish only when $\widehat{H}(q)$ possesses a zero-energy eigenvalue, since $\mathrm{Det}(U)=\pm1$, and $\mathrm{Pf}[U\widehat{H}(q)U^\dagger]=\mathrm{Det}(A(q))$. One may therefore define a complex function $z(q)=\exp[i\theta(q)]=\mathrm{Det}(A(q))/|\mathrm{Det}(A(q))|$, which has unit modulus $|z(q)|=1$, following Ref.~\cite{PhysRevB.99.035312}. This allows the periodic disorder invariant to be written as
\begin{equation}
    \nu = \frac{-i}{\pi}\int_{q=0}^{q=\pi} \frac{dz(q)}{z(q)}.
\end{equation}

\noindent Proceeding as in the clean system\cite{PhysRevB.99.035312,MoorePRB2007}, the periodic disorder invariant may be related to the Pfaffian invariant as

\balance
\begin{widetext}
\begin{eqnarray}   
    \mathrm{sgn}[\mathrm{Pf}({\widehat{H}(q=0)})\mathrm{Pf}(\widehat{H}(q=\pi))] \equiv \mathrm{sgn}\left[\frac{\mathrm{Pf}({\widehat{H}(q=\pi)})}{\mathrm{Pf}(\widehat{H}(q=0))}\right]=\mathrm{sgn}\left[\frac{\mathrm{Det}({A(q=\pi)})}{\mathrm{Det}(A(q=0))}\right]
    =\frac{z(q=\pi)}{z(q=0)}=e^{i\pi\nu} = (-1)^\nu.
\end{eqnarray}
\end{widetext}
Thus the parity of the PDI coincides with the sign of the superlattice Pfaffian invariant. The equivalence between the parity of the periodic disorder invariant and the sign of the superlattice Pfaffian invariant therefore establishes the latter as a well-defined $\mathbb{Z}_2$ topological invariant for spatially inhomogeneous systems. We now demonstrate how this superlattice formulation is connected to the twisted-boundary Pfaffian invariant introduced earlier.

Thus far the superlattice Pfaffian invariant, in Eq.~\ref{eq:sup_latt_Pf}, has been formulated under periodic boundary conditions, which allow the definition of the superlattice Bloch momenta. To connect this formulation with the twisted-boundary Pfaffian invariant introduced earlier for finite nanowires (refer to Sec.~\ref{sec:relation_bw_ms_rs}), we now consider threading a magnetic flux ($\Phi$) through the superconducting ring formed by coupling the last microscopic site ($j=L-1$, where $L=N_c l$) to the first site ($j=0$) of the superlattice. The presence of the magnetic flux introduces a Peierls phase in the boundary hopping. This introduces a term in the superlattice Hamiltonian, $\mathrm{H}_{\phi}$, representing the flux-induced phase-dependent hopping between the ends of the first ($n=0,j=0$) and last supercells ($n=N_c-1,j=L-1$)

\balance
\begin{widetext}
    \begin{equation}
    \widetilde{H} = \frac{1}{2}\sum_n\left[\Psi^\dagger_n\mathrm{H_0}\Psi_n + \Psi^\dagger_{n+1}\mathrm{H_1}\Psi_n+\Psi^\dagger_{n}\mathrm{H^\dagger_1}\Psi_{n+1}\right] + \Psi^\dagger_{N_c-1}\mathrm{H_\phi}\Psi_0+\Psi^\dagger_{0}\mathrm{H^\dagger_\phi}\Psi_{N_c-1}
    \label{eq:suplatt_no_phase}
\end{equation}
\end{widetext}

\noindent where $\phi=\pi\Phi/\Phi_0$ denotes the flux-induced phase twist and $\Phi_0=h/2e$ is the flux quanta. As in Sec.~\ref{sec:relation_bw_ms_rs}, this phase may be redistributed into the bulk of the superlattice through a gauge transformation. To account for this phase distributed throughout the bulk of the superlattice, the Hamiltonian takes the form

\balance
\begin{widetext}
\begin{equation}
    \widetilde{H} = \frac{1}{2}\sum_n\left[\Psi^\dagger_n\mathrm{H_0}\Psi_n + e^{-i\phi/N_c}\Psi^\dagger_{n+1}\mathrm{H_1}\Psi_n+e^{i\phi/N_c}\Psi^\dagger_{n}\mathrm{H^\dagger_1}\Psi_{n+1}\right] + e^{-i\phi(N_c-1)/N_c}\Psi^\dagger_{N_c-1}\mathrm{H_1}\Psi_0+e^{i\phi(N_c-1)/N_c}\Psi^\dagger_{0}\mathrm{H^\dagger_1}\Psi_{N_c-1}.
\end{equation}
\end{widetext}

\noindent As a consequence, the allowed superlattice Bloch momenta are modified according to

\begin{equation}
    q = \frac{2\pi n + \phi}{N_c}, \qquad n=0,1,\ldots,N_c-1
    \label{eq:bloch_momenta_phase}
\end{equation}

This modification does not alter the Bloch representation of the superlattice Hamiltonian given in Eq.~\ref{hatH}, since the flux-induced phase is incorporated directly through the shifted quantization of the Bloch momentum in Eq.~\ref{eq:bloch_momenta_phase}. One may therefore follow the same boundary-twist argument as in Sec.~\ref{sec:relation_bw_ms_rs} to determine how the parity of $N_c$, together with the special values $\phi=0$ and $\phi=\pi$, fixes the sign of the Pfaffian under periodic boundary conditions,

\balance
\begin{widetext}
\begin{equation}
    \mathrm{sgn[Pf[\widetilde{H}(\phi=0)]]}=\begin{cases}
        \mathrm{sgn[Pf}[\widetilde{H}(q=0)]\times \mathrm{Pf}[\widetilde{H}(q=\pi)]], & \text{for even $N_c$} \\
        \mathrm{sgn[Pf}[\widetilde{H}(q=0)]], & \text{for odd $N_c$}
    \end{cases}
\end{equation}
\end{widetext}

\noindent while the Pfaffian under anti-periodic boundary conditions takes the form
\begin{equation}
    \mathrm{sgn[Pf}[\widetilde{H}(\phi=\pi)]]=\begin{cases}
        \mathrm{+1}, & \text{for even $N_c$} \\
        \mathrm{sgn[Pf}[\widetilde{H}(q=\pi)]], & \text{for odd $N_c$}
    \end{cases}.
\end{equation}

\noindent Consequently, the total Pfaffian $\mathrm{sgn[Pf}[\widetilde{H}(\phi=0)]\mathrm{Pf}[\widetilde{H}(\phi=\pi)]]$ is determined entirely by the particle–hole symmetric points $q=0,\pi$ of the superlattice Brillouin zone, irrespective of whether $N_c$ is even or odd. This establishes that the topological characterization obtained within the superlattice Bloch momentum formulation $\widehat{H}(q)$ is equivalent to the real-space Pfaffian invariant defined through periodic and anti-periodic boundary conditions in the superlattice formulation $\widetilde{H}(\phi)$,

\begin{eqnarray}
    \mathrm{sgn[Pf}(\widehat{H}(q=0))\mathrm{Pf}(\widehat{H}(q=\pi))] &\equiv\nonumber \\
    \mathrm{sgn[Pf}(\widetilde{H}(\phi=0))\mathrm{Pf}(\widetilde{H}(\phi=\pi))]
\end{eqnarray}

A particularly transparent limit is obtained when $N_c=1$, for which the superlattice consists of a single supercell. In this case the superlattice Bloch momentum reduces directly to the flux-induced phase twist $q=\phi$. In this limit $\widehat{H}(q)$ in Eq.~\ref{hatH} becomes identical to the Hamiltonian of a single supercell $\mathcal{H}(\phi)$ given in Eq.~\ref{eq:single_chain_H} with twisted boundary conditions. Consequently, the Pfaffian invariant formulated in superlattice Bloch momentum space in Eq.~\ref{eq:sup_latt_Pf} coincides with the real-space Pfaffian invariant in Eq.~\ref{eq:real_sp_Pfaff} evaluated for a unit supercell, demonstrating that the twisted-boundary Pfaffian invariant remains a valid topological invariant even in the presence of disorder.

\begin{eqnarray}
    \mathrm{sgn}[\mathrm{Pf}(\widehat{H}(q=0))\mathrm{Pf}(\widehat{H}(q=\pi))] &\equiv& \nonumber\\ 
    \mathrm{sgn}\left[\mathrm{Pf({\mathcal{H}(\phi=0)})}\mathrm{Pf(\mathrm{\mathcal{H}(\phi=\pi)})}\right]
\end{eqnarray}

\noindent In summary, we have shown that the Pfaffian invariant formulated in terms of the superlattice Bloch momenta $q=0,\pi$ in Eq.~\ref{eq:sup_latt_Pf} is equivalent to the parity of the periodic disorder invariant in Eq.~\ref{nu}. By introducing a flux-induced phase twist in the superlattice boundary hopping, we further established that this superlattice Pfaffian invariant reduces to the twisted-boundary Pfaffian invariant, in Eq.~\ref{eq:real_sp_Pfaff}, of a finite SM-SC nanowire in the limit $N_c=1$. These results demonstrate that the twisted-boundary Pfaffian invariant defined through periodic and anti-periodic boundary conditions remains a well-defined $\mathbb{Z}_2$ topological invariant even in the presence of disorder.

\section{Relationship between the Pfaffian of the Hamiltonian and the Ground state Fermion Parity}
\label{sec:proof_fermion_parity}

In the previous section we demonstrated that several seemingly different topological invariants—the momentum-space Pfaffian invariant, the twisted-boundary Pfaffian invariant, and the superlattice formulation related to the PDI—are equivalent diagnostics of topology in SM–SC nanowires. We now establish the physical meaning of the Pfaffian itself by showing that the sign of the Pfaffian of a quadratic Hamiltonian determines the fermion parity of its ground state.

Any quadratic Hamiltonian may be written in the Majorana form \cite[Eq.~3]{Kitaev2001}
\begin{equation} 
H = \frac{i}{4} \sum_{l,m} A_{lm} \gamma_l \gamma_m.
\end{equation}
Here $\gamma_i$ are Majorana operators and $A$ is a real skew-symmetric matrix satisfying $A_{lm}^*=A_{lm}=-A_{ml}$. In this representation, the Hamiltonian is fully characterized by the antisymmetric matrix $A$, and the phrase ``Pfaffian of the Hamiltonian'' refers to $\mathrm{Pf}(A)$.

To define $\mathrm{Pf}(A)$ explicitly, we introduce the symmetric group $S_{2N}$, the group of all permutations of $(1,2,\ldots,2N)$. This group is generated by transpositions $\sigma_i=(i,i+1)$ for $1\le i\le 2N-1$, so that every $\sigma\in S_{2N}$ can be written as a product of transpositions $\sigma=\prod \sigma_j$. The sign of a permutation $\sigma$, denoted $\mathrm{sgn}(\sigma)$, is defined as
\begin{equation}
\operatorname{sgn}(\sigma) =
\begin{cases}
+1 & \text{even number of transpositions},\\
-1 & \text{odd number of transpositions}.
\end{cases}
\end{equation}
\noindent By even (odd) we mean that $\sigma$ can be expressed as $\sigma=\prod_{j=1}^{2m}\sigma_j$ ($\sigma=\prod_{j=1}^{2m-1}\sigma_j$).
\vspace{0.1cm}
The Pfaffian of a real skew-symmetric matrix $A$ is given by
\begin{equation} 
\text{Pf} (A) = \frac{1}{2^NN!} \sum_{\sigma \in S_{2N}} \text{sgn}(\sigma) A_{\sigma(1),\sigma(2)} \cdots A_{\sigma(2N-1),\sigma(2N)}, 
\end{equation}
\noindent and satisfies $(\text{Pf}(A))^2=\det(A)$.

\noindent Consider a matrix of the form 
\begin{equation}
M = \bigoplus_{i=1}^{N} \begin{pmatrix}
    0 & \epsilon_i \\
    -\epsilon_i & 0
\end{pmatrix}.
\end{equation}
\noindent Since the Pfaffian is multiplicative under direct sums, it follows that
\begin{equation}
\text{Pf}\lrp{\bigoplus_{i=1}^{N} \begin{pmatrix}
    0 & \epsilon_i \\
    -\epsilon_i & 0
\end{pmatrix}} = \prod_{i=1}^{N} \text{Pf} \begin{pmatrix}
    0 & \epsilon_i \\
    -\epsilon_i & 0 
\end{pmatrix} = \prod_{i=1}^{N} \epsilon_i.
\end{equation}

\noindent Only permutations belonging to the subgroup $P=\langle (2i-1,2i), (2j-1,2j+1)(2j,2j+2)\rangle$ contribute nonvanishing terms to the summation. This subgroup has size $2^NN!$. The Pfaffian of the direct sum may therefore be written as
$$\text{Pf}(M)=\sum_{\sigma \in P}\frac{\text{sgn}(\sigma)}{2^NN!} \lrp{\pm \prod_{i=1}^N\epsilon_i}.$$
If $\text{sgn}(\sigma)=-1$, the negative sign must be chosen from the $\pm$ factor arising from the definition of $M_{2i-1}M_{2i}$. Thus,
\begin{align}
\text{Pf}(M) &= \sum_{\sigma \in P}\frac{\pm 1}{2^NN!} \lrp{ \pm \prod_{i=1}^N\epsilon_i} \\
&= \prod_{i=1}^N \epsilon_i.
\end{align}
\noindent This establishes that $\text{Pf}(M)=\displaystyle\prod_i \epsilon_i \geq 0$.

\noindent For the $2N\times 2N$ skew-symmetric Majorana matrix $A$, one may block-diagonalize it by an orthogonal transformation $W$, yielding $WAW^T$. It is well known that $\text{Pf}(WAW^T)=\det(W)\,\text{Pf}(A)$ \cite[Eq.~21]{Kitaev2001}. If $W$ can be chosen such that
\begin{equation}
WAW^T=\bigoplus_{i=1}^{N} \begin{pmatrix}
    0 & \epsilon_i \\
    -\epsilon_i & 0
\end{pmatrix},
\end{equation}
\noindent then
\begin{equation}\label{wawequality}
\text{Pf}(WAW^T) = \text{Pf}(A)\,\det(W) \geq 0,
\end{equation}
\noindent and therefore
\begin{equation}
    \text{sgn}(\text{Pf}(A))=\text{sgn}(\det(W)).
\end{equation}

\noindent Take $W$ to be the $2N\times 2N$ real orthogonal matrix whose rows are the eigenvectors of $A$. Define new operators $\gamma_i'$ by
$$ \gamma_i' = \sum_{j=1}^{2N} W_{ij} \gamma_j. $$
Since $W \in O(2N)$, it follows that $\det(W)=\pm1$.

\noindent The row-orthogonality condition $WW^{T} = I$ reads
\begin{equation}\label{row_ortho}
\sum_{j=1}^{2N} W_{ij}\, W_{kj} = \delta_{ik},
\qquad i,k = 1,\dots,2N .
\end{equation}

The fermion parity operator in the $\gamma_i$ basis is given by $P=\displaystyle\prod_{i=1}^{N} (i\gamma_{2i-1}\gamma_{2i})$, and it anticommutes with every Majorana operator, $\{P, \gamma_j\}=0$.

\noindent In the transformed basis, we have
\begin{align}
P'=\prod_{i=1}^{N}(i\gamma_{2i-1}'\gamma_{2i}')&=i^N\gamma_1'\gamma_2'\cdots\gamma_{2N}'\\
&=i^N \prod_{j=1}^{2N}\gamma_j'\\ 
&= i^N\prod_{j=1}^{2N} \lrp{ \sum_{k=1}^{2N} W_{jk}\gamma_k }.
\end{align}

\noindent Expanding this expression yields
\begin{equation}\label{iniparity}
P' = i^N \sum_{k_1,\cdots,k_{2N}} \lrp{ \prod_{j=1}^{2N} W_{jk_j} } \gamma_{k_1} \cdots \gamma_{k_{2N}}.
\end{equation}
In the summation above, terms with repeated indices $k_m=k_n=f$ may be reordered using the Majorana commutation relations so that identical operators are brought together,
\begin{align}
\gamma_{k_1}&\cdots\gamma_{k_{r-1}}\,
\underline{\gamma_{f}}\,(\text{other }\gamma\text{'s})\,
\underline{\gamma_{f}}\,
\gamma_{k_{s+1}}\cdots\gamma_{k_{2N}}
\\ &= \pm \,(\text{product of $2N-2$ } \gamma\text{'s}).
\end{align}
In this manner, any term containing repeated Majorana operators may be reduced to one involving fewer operators. If a term contains an odd number of a given $\gamma$ operator, one may choose pairs to reduce, leaving a term involving distinct operators. A general reduced term takes the form
\begin{align}
\lrp{ i^N W_{m_1f_1}W_{n_1f_1}\cdots W_{m_lf_l}W_{n_lf_l} }\prod_{\substack{j=1 \\ j\neq m_i,n_i}}^{2N-2l}  \lrp{W_{jk_j} \gamma_{k_1} \cdots \gamma_{k_{2N}}}.
\end{align}

Discarding the constant and summing all terms with $k_{m_i}=k_{n_i}=f_i$, we obtain
\begin{equation}\label{eq:xprod}
X.\lrp{\sum_{\substack{k_1,\cdots k_{m_i-1},k_{m_i+1}, \\ \cdots k_{n_i-1},k_{n_i+1},k_{2N}}} \lrp{ \prod_{\substack{j=1 \\ j\neq m_i,n_i}}^{2N-2l} W_{jk_j} } \gamma_{k_1} \cdots \gamma_{k_{2N}}}.
\end{equation}
\noindent where
\begin{equation}
    X = W_{m_1f_1}W_{n_1f_1}\cdots W_{m_lf_l}W_{n_lf_l}.
\end{equation}

The term multiplying $X$ in Eq.~\ref{eq:xprod} (denoted $D$) is identical for all sums with repeated operators at the same pair of indices. Summing over these contributions yields
\begin{equation}
\sum_{f_i}^{2N}({W_{m_1f_1}W_{n_1f_1})\cdots (W_{m_lf_l}W_{n_lf_l}})D.
\end{equation}
This expression vanishes upon summation over $f_i$ since $W$ is an orthogonal matrix. Eq.~\ref{row_ortho} has been used at this step, and no assumption beyond $m \neq n$ is required. Consequently, only terms with distinct indices survive in the summation.

When all indices are distinct,
\begin{equation} \label{reorder}
(k_1,\ldots,k_{2N}) = (\sigma(1),\ldots,\sigma(2N)),
\end{equation}
\noindent the product of Majorana operators may be reordered as
\begin{equation}
\gamma_{k_1}\cdots\gamma_{k_{2N}}
= \operatorname{sgn}(\sigma)\,\gamma_1\gamma_2\cdots\gamma_{2N}.
\end{equation}
Combining Eq.~\ref{iniparity} and Eq.~\ref{reorder}, and restricting to terms with distinct $k_j$, one obtains
\begin{align}
    P' &= i^N \lrp{ \sum_{\sigma \in S_{2N}} \text{sgn}(\sigma) \prod_{j=1}^{2N} W_{j\sigma(j)}} \gamma_1 \cdots \gamma_{2N} \\ 
    &= i^N \det(W) \gamma_1 \cdots \gamma_{2N}. 
\end{align}
Thus, $P = \det(W)^{-1} P'$. By convention, we choose $\bra{G}P' \ket{G}=1$. Combining this with $\text{sgn}(\det(W))=\text{sgn}(\text{Pf}(A))$, it follows that
\begin{equation}\label{signequalsparity}
\bra{G}P \ket{G}=\text{sgn}(\text{Pf}(A)) \cdot \bra{G}P' \ket{G}= \text{sgn}(\text{Pf}(A)).
\end{equation}
Thus, the sign of the Pfaffian of a quadratic Hamiltonian encodes the fermion parity of its ground state, and $\mathrm{sgn}[\mathrm{Pf}(A)]$ may be viewed as a ground-state fermion parity indicator for quadratic superconducting systems. This result establishes a direct physical interpretation of the Pfaffian invariant as a ground-state fermion parity indicator. In the following section we illustrate these relations numerically by examining flux-induced level crossings and parity switches in finite and disordered nanowire systems.

\section{Numerical Results}
\label{sec:numerical_results}

\begin{figure}[t]
    \centering
    \includegraphics[width=\linewidth]{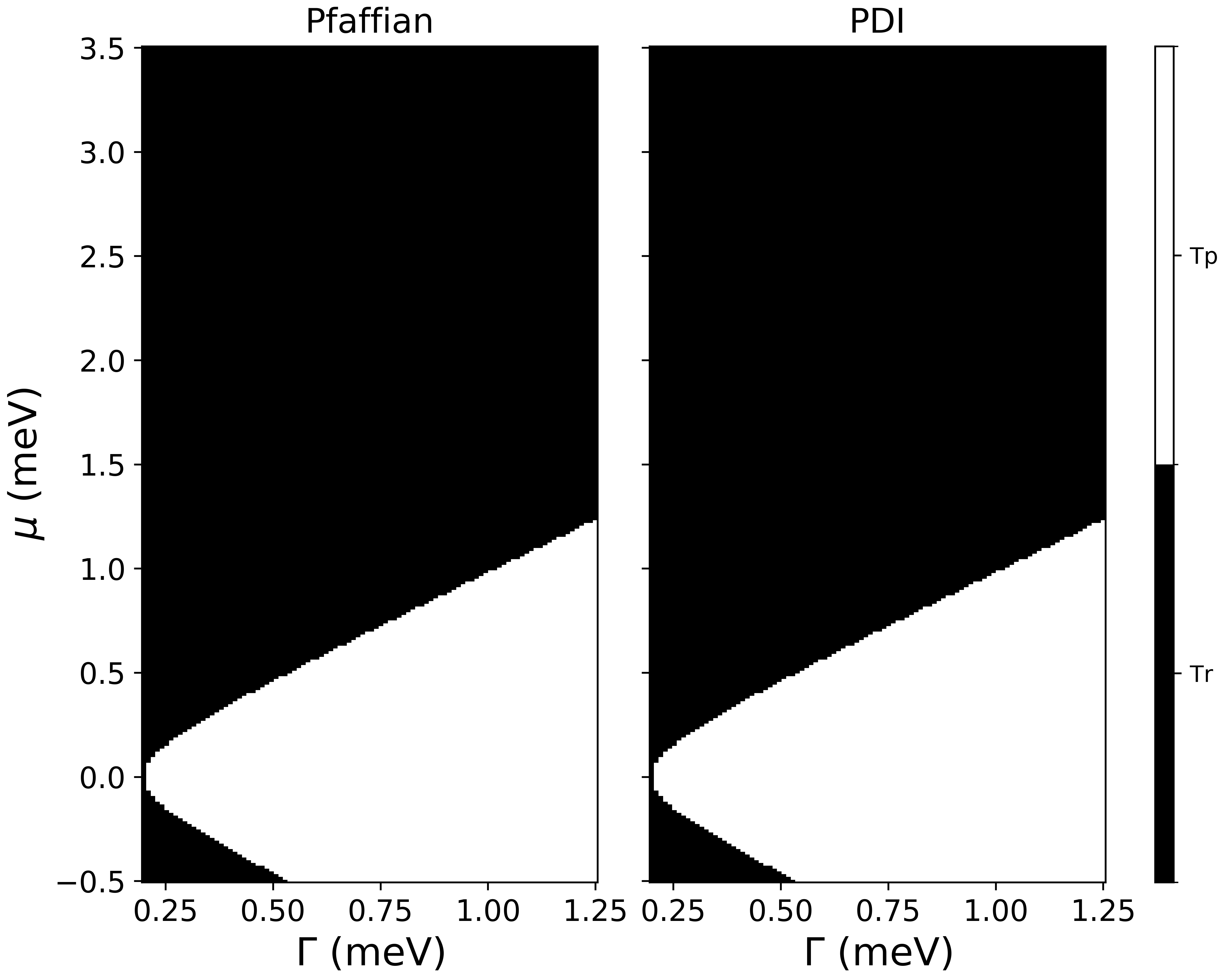}
    \caption{Invariant maps for a clean $3\,\,\mu$m long wire. The left panel shows the real space twisted-boundary Pfaffian invariant (refer to Sec.~\ref{sec:relation_bw_ms_rs}) and the right panel shows the corresponding PDI (refer to Sec.~\ref{sec:PDI}). The black regions indicate topologically trivial region ($Tr$) where the Pfaffian invariant has a positive sign while PDI has a value of 0 and the white regions are the topologically non-trivial ($Tp$) regimes where the Pfaffian invariant has a negative sign and PDI is 1.}
    \label{fig:Pf_PDI_clean_map}
\end{figure}

We now present numerical results illustrating the correspondence between the Pfaffian-based invariants, ground state fermion parity switched and flux-induced low energy level crossings. We analyze the magnetic flux $\Phi$ dependence of the lowest BdG eigenvalues for a nanowire closed into a ring geometry. In this configuration, a magnetic flux threading the loop induces a phase twist ($\phi=\pi\Phi/\Phi_0$ where $\Phi_0=h/2e$ is the flux quantum) across the ends of the wire, providing a direct probe of fermion parity switching and boundary condition sensitivity \cite{PRBSau2025,ArXivStanescu2025,MSRMicrosoft2024}.

As a benchmark, we first consider a clean nanowire system. The wire has length $3~\mu\mathrm{m}$ (300 lattice sites) with lattice constant $a = 10~\mathrm{nm}$. The effective tight binding parameters include nearest neighbor hopping $t = 16.56~\mathrm{meV}$, chemical potential $\mu$, Rashba spin-orbit coupling strength $\alpha = 1.4~\mathrm{meV}$, and Zeeman field $\Gamma$. Proximity induced superconductivity is characterized by a pairing potential $\Delta = 0.2~\mathrm{meV}$, placing the system in the weak coupling regime between the semiconductor and the superconductor.

Fig.~\ref{fig:Pf_PDI_clean_map} compares the real space Pfaffian invariant and the periodic disorder invariant (PDI) for the clean system. The two phase diagrams are indistinguishable across the entire $(\mu,\Gamma)$ parameter space, establishing a direct numerical correspondence between the two formulations despite their distinct computational implementations \cite{WimmerACM2012,eissele2025topological}.

To connect this invariant-based characterization with the low energy spectral properties of the system, we examine the behavior of the lowest BdG eigenvalues as a function of flux. In particular, we focus on the occurrence of near zero energy crossings, which signal fermion parity switching \cite{PRBSau2025} and provide a physical manifestation of the topological phase. Such a crossing between $\phi=0$ and $\phi=\pi$ implies that each parity-resolved energy branch acquires an $h/e$ periodicity in flux, rather than the conventional $h/2e$ periodicity, reflecting the $4\pi$-periodic Josephson effect characteristic of topological superconductivity\cite{Kitaev2001}.

Representative points are selected from distinct regions of the phase diagram: (i) a topologically trivial point at $(\mu=1.0~\mathrm{meV},\,\Gamma=0.5~\mathrm{meV})$ and (ii) a point within a topological island at $(\mu=0.0~\mathrm{meV},\,\Gamma=0.5~\mathrm{meV})$. These choices allow a direct comparison between the flux dependent spectral response in trivial and topological regimes.

\begin{figure}[h]
    \centering
    \includegraphics[width=\linewidth]{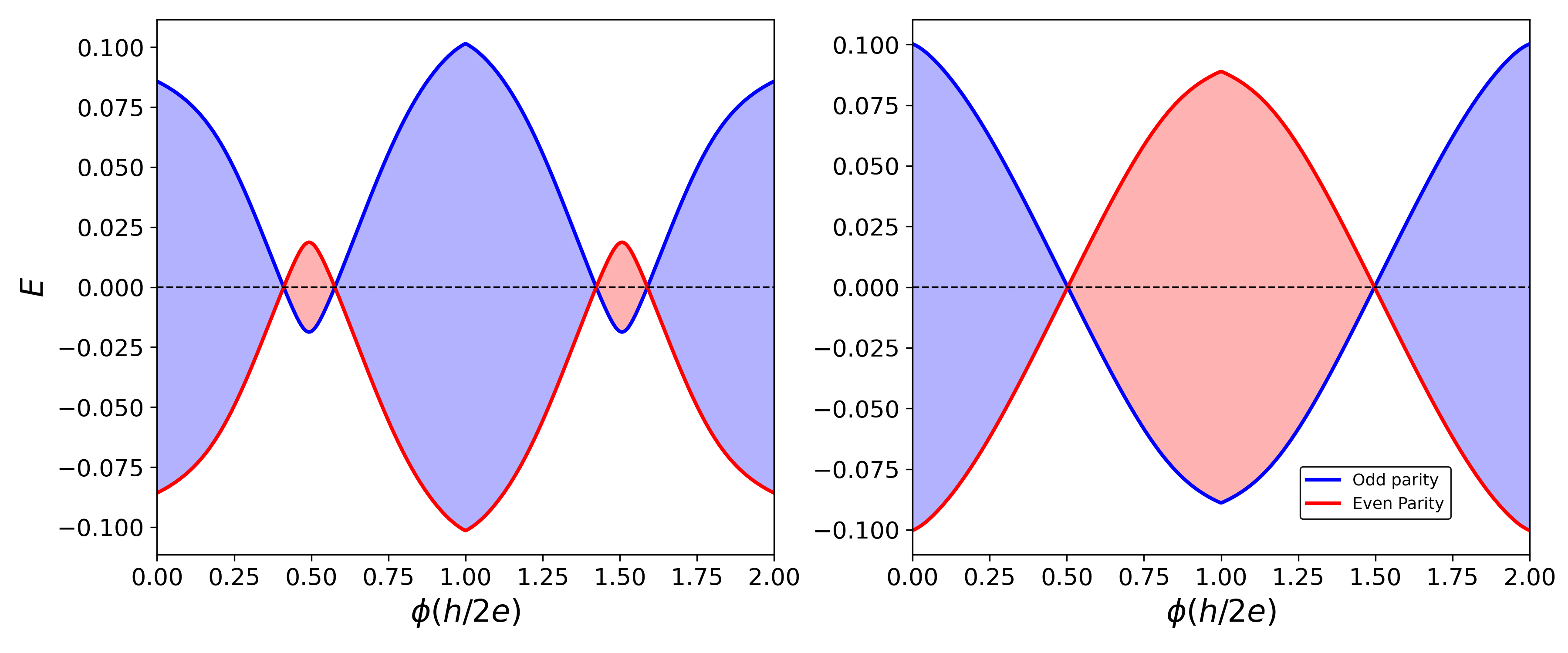}
    \caption{Flux dependent energy spectrum for a clean $3\,\,\mu$m long wire. The left panel shows the variation of 2 lowest BdG energy eigenvalues for the trivial region point ($\mu=1.0$~meV,$\Gamma=0.5$~meV) and the right panel shows the behavior of the 2 energy curves for topological regime ($\mu=0.0$~meV,$\Gamma=0.5$~meV). The blue curve corresponds to odd parity while the red curve corresponds to even parity.}
    \label{fig:Evsphi_clean}
\end{figure}

\begin{figure*}[ht]
    \centering
    \includegraphics[width=\linewidth]{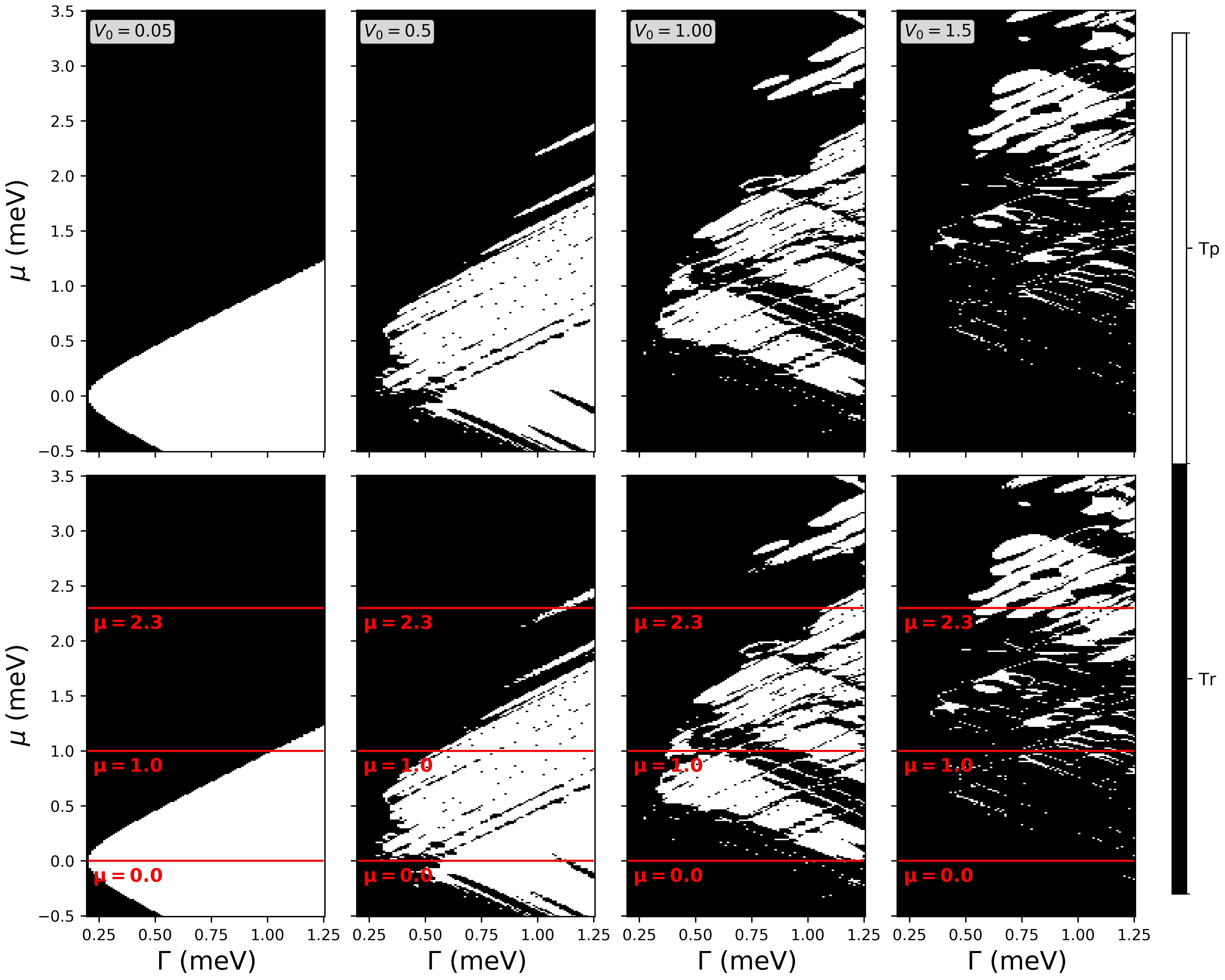}
    \caption{Top row: Real space twsited-boundary Pfaffian invariant (refer to Sec.~\ref{sec:relation_bw_ms_rs}) maps for varying chemical potential $\mu$ on y-axis and Zeeman field $\Gamma$ on x-axis for increasing disorder amplitude $V_0=0.05, 0.5, 1.0$ and $1.5$ meV. Bottom row: Corresponding PDI maps (refer to Sec.~\ref{sec:PDI}) for the same set of parameters.}
    \label{fig:Pf_PDI_dis_map}
\end{figure*}

The flux dependent low energy spectrum is shown in Fig.~\ref{fig:Evsphi_clean}, where the evolution of the lowest BdG eigenvalues is plotted as a function of the phase $\phi$ induced by the magnetic flux. A key diagnostic of topology is the parity of the number of zero energy crossings between $\Phi=0$ and $\Phi=h/2e$, which distinguishes topologically trivial and nontrivial regimes.

In the left panel of Fig.~\ref{fig:Evsphi_clean}, corresponding to the trivial parameter point, the spectrum shows two $h/2e$ periodic low energy levels. The spectrum exhibits a pair of narrow double crossings, resulting in an even number of crossings over the interval. As a consequence, there is no net change in the sign of either the even- or odd-parity energy branches between $\Phi=0$ and $\Phi=h/2e$, and the ground state fermion parity remains unchanged. This corresponds to the standard $2\pi$ Josephson effect.

By contrast, the right panel displays the spectrum for a point within the topological region, where there is a true zero energy crossing occurs between $\Phi=0$ and $\Phi=h/2e$. The presence of a true crossing in this interval implies that each parity-resolved energy branch becomes $h/e$ periodic in flux, rather than $h/2e$ periodic. Correspondingly, both the even- and odd-parity branches change sign over the interval, resulting in a switch of the ground state fermion parity at $\Phi=h/2e$. This behavior is the spectral manifestation of the $4\pi$-periodic Josephson effect characteristic of the topological phase.

These flux induced parity switches provide a direct spectral signature of the topological invariant, establishing that both the real space Pfaffian invariant and the periodic disorder invariant capture fermion parity switching in the ground state. Having benchmarked this correspondence in the clean system, we now turn to the disordered case with finite disorder strength.

Disorder in the finite semiconductor wire is modeled by a random onsite potential $V_{\mathrm{dis}}(i)$ satisfying $\langle V_{\mathrm{dis}} \rangle = 0$ and $\langle V_{\mathrm{dis}}^2 \rangle = V_0^2$. In the weak coupling regime, fluctuations in the SM chemical potential are only weakly screened by the parent superconductor, leading to an enhanced sensitivity of the topological phase structure to disorder\cite{PhysRevB.110.115436}. As a result, disorder strongly influences the distribution and fragmentation of topological regions (``topological islands'') in the $\mu$-$\Gamma$ parameter space making this regime serve a true test of equivalence between real space Pfaffian invariant and PDI.

Fig.~\ref{fig:Pf_PDI_dis_map} presents a comparison of the two real space formulations of the topological invariant in the presence of disorder, with the top row showing the twisted-boundary Pfaffian invariant maps and the bottom row displaying the corresponding PDI maps. Across the full range of disorder strengths considered, from weak disorder ($V_0=0.05~\mathrm{meV}$) to strong disorder ($V_0=1.5~\mathrm{meV}$), the two sets of phase diagrams remain visually indistinguishable.

A quantitative, pixel by pixel comparison reveals discrepancies in only a small number of points of order $10^1$ out of $3\times10^4$—localized near the boundaries of the topological regions where the system transitions between trivial and nontrivial phases. These differences arise close to gap closing points, where the numerical evaluation of the invariants is particularly sensitive, and can therefore be attributed to the distinct numerical implementations rather than to any physical inconsistency.

To examine whether this agreement between the Pfaffian and PDI maps is reflected in the low energy spectral response, we consider horizontal cuts through the phase diagrams at fixed chemical potentials $\mu = 0.0$, $1.0$, and $2.3~\mathrm{meV}$, indicated by the thick red lines in the bottom row of Fig.~\ref{fig:Pf_PDI_dis_map}. Along these cuts, we analyze whether the regions identified as topological by the invariants exhibit fermion parity switching through flux induced level crossings.

To this end, we define a fermion parity switch indicator (FPSI) $\mathcal{F}$, which takes binary values according to
\begin{equation}
    \mathcal{F} = 
    \begin{cases}
        1, & \Delta E_g(\Phi=0)\,\Delta E_g(\Phi=h/2e) < 0, \\
        0, & \Delta E_g(\Phi=0)\,\Delta E_g(\Phi=h/2e) > 0 ,
    \end{cases}
    \label{eq:FPSI}
\end{equation}
where $\Delta E_g(\phi) = E_{\mathrm{even}}(\Phi) - E_{\mathrm{odd}}(\Phi)$ denotes the energy difference between the lowest energy states in the even and odd fermion parity sectors at a given flux. A negative value of the product $\Delta E_g(\Phi=0)\,\Delta E_g(\Phi=h/2e)$ indicates a reversal in the relative ordering of the parity resolved ground states between $\Phi=0$ and $\Phi=h/2e$, corresponding to a true fermion parity switch. At a fixed value of $\phi$, the ground state fermion parity is determined by the lower of the two sector resolved energies $E_{\mathrm{even}}(\phi)$ and $E_{\mathrm{odd}}(\phi)$. Equivalently,
\begin{equation}
P_{\mathrm{GS}}(\phi)=
\begin{cases}
+1, & \Delta E_g(\phi)<0 \quad \big(E_{\mathrm{even}}<E_{\mathrm{odd}}\big),\\
-1, & \Delta E_g(\phi)>0 \quad \big(E_{\mathrm{odd}}<E_{\mathrm{even}}\big).
\end{cases}
\end{equation}
Thus, a change in the sign of $\Delta E_g(\phi)$ corresponds to an exchange in the ordering of the even- and odd-parity ground states, i.e., a fermion-parity switch of the many-body ground state. Accordingly, the FPSI $\mathcal{F}$ equals unity when a fermion parity switch occurs and vanishes otherwise, either due to an even number of zero energy crossings, as observed in the trivial regime of the clean system in Fig.~\ref{fig:Evsphi_clean}, or due to avoided crossings. In a superconducting BdG system the fermion parity of the ground state changes whenever a quasiparticle level crosses zero energy \cite{Kitaev2001,tewari2012topologicaalinvariant}. Such crossings correspond to a closing and reopening of the lowest excitation gap $\Delta E_g$, and therefore a change in the sign of $\Delta E_g$ signals a fermion-parity switch of the ground state. The FPSI $\mathcal{F}$ therefore serves as a measure of fermion-parity switching, since it detects these sign changes in $\Delta E_g$. In this sense, the FPSI captures the correspondence between the real space twisted-boundary Pfaffian invariant (Sec.~\ref{sec:relation_bw_ms_rs}, Eq.~\ref{eq:real_sp_Pfaff}) and the fermion parity of the ground state established in Sec.~\ref{sec:proof_fermion_parity}. The numerical results presented below demonstrate that changes in the FPSI occur precisely at the points where the topological invariant changes sign.

\begin{figure}
    \centering
    \includegraphics[width=\linewidth]{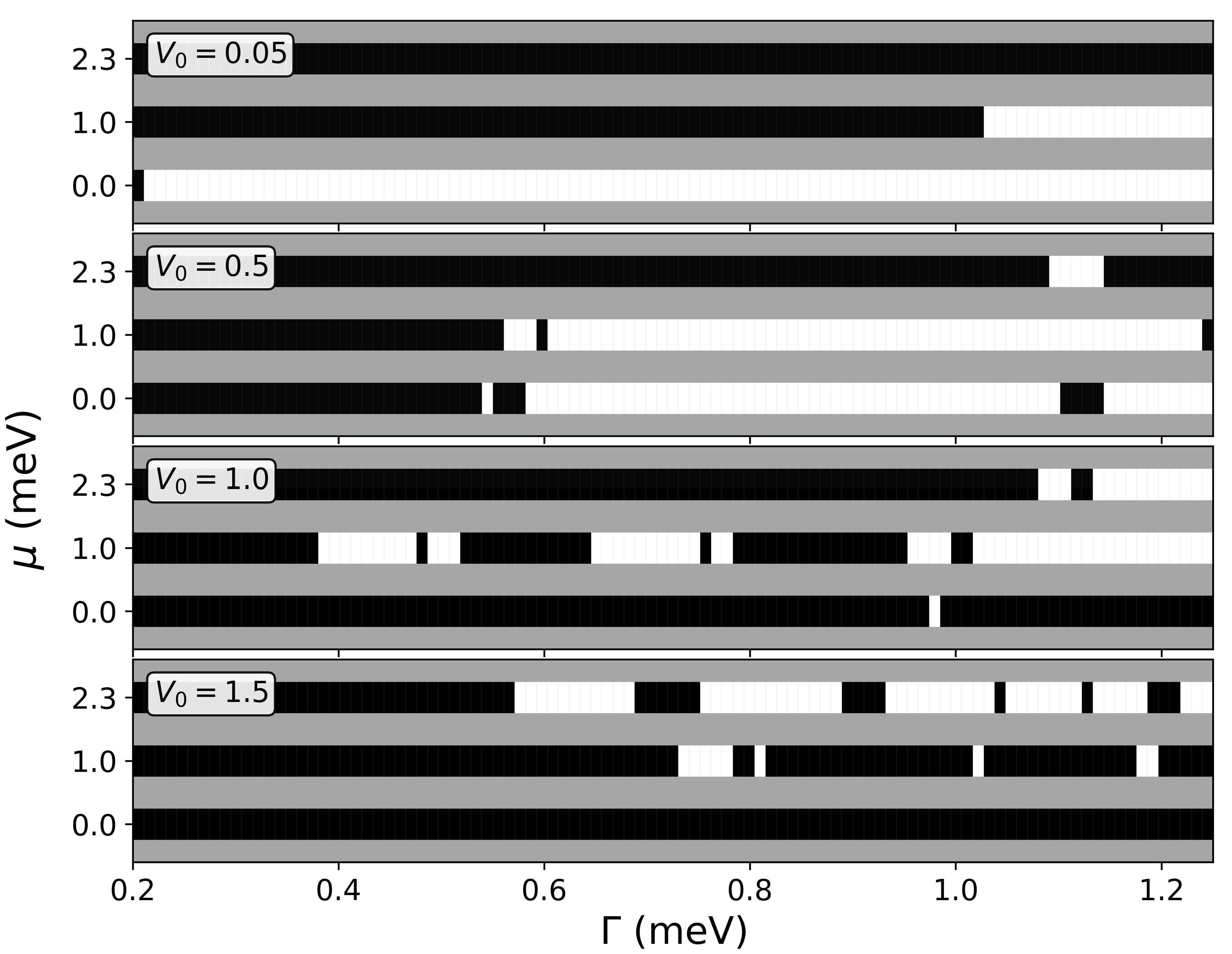}
    \caption{Fermion parity switch indicator $\mathcal{F}$ (refer to Eq.~\ref{eq:FPSI}) as a function of Zeeman field for different chemical potential values, $\mu=[0.0,1.0,2.3]$~meV for increasing disorder in the nanowire. The top panel corresponds to weak disorder limit $V_0=0.05$~meV and moving down towards the bottom panel representing the high disorder limit $V_0=1.5$~meV. The ribbon cuts in each panel show the value of the fermion parity switch indicator as black corresponding to $\mathcal{F}=0$ (no fermion parity switch) and white corresponding to $\mathcal{F}=1$ (fermion parity switch taking place).}
    \label{fig:H_cuts_mu}
\end{figure}

The horizontal cuts through the disordered phase diagrams shown in Fig.~\ref{fig:Pf_PDI_dis_map} are analyzed using the fermion parity switch indicator $P$, with the corresponding results displayed in Fig.~\ref{fig:H_cuts_mu}. The panels are arranged in order of increasing disorder strength, from weak disorder in the top row ($V_0=0.05~\mathrm{meV}$) to strong disorder in the bottom row ($V_0=1.5~\mathrm{meV}$). For each fixed chemical potential, the FPSI results may be directly compared with the corresponding horizontal cuts in the Pfaffian and PDI maps. Regions identified as topological by the invariants (white) consistently coincide with values of $\mathcal{F}=1$ (also indicated by white in Fig.~\ref{fig:H_cuts_mu}), indicating the presence of a fermion parity switch, while trivial regions (black) correspond to $\mathcal{F}=0$ (indicated by black in Fig.~\ref{fig:H_cuts_mu}). The Zeeman field values at which $\mathcal{F}$ changes from $0$ to $1$ (or vice versa) align with the boundaries of the topological regions in the invariant maps. This correspondence is evident for all chemical potential cuts and remains robust as the distribution of topological regions evolves with increasing disorder. For brevity, we focus on the $\mu=2.3~\mathrm{meV}$ cut, which provides a representative illustration. In the weakly disordered regime ($V_0=0.05~\mathrm{meV}$), this cut lies entirely within the trivial region of the phase diagram, and the FPSI remains zero across the full range of Zeeman field values. At intermediate disorder strength ($V_0=0.5~\mathrm{meV}$), a localized region with $\mathcal{F}=1$ emerges, reflecting the intersection of the horizontal cut with a small topological island in the invariant maps. As the disorder strength increases to $V_0=1.0~\mathrm{meV}$, this topological region expands, with the FPSI indicating a fermion parity switch beginning at approximately $\Gamma \simeq 1.06~\mathrm{meV}$, in agreement with the corresponding phase boundary in the invariant maps. In the strongly disordered regime ($V_0=1.5~\mathrm{meV}$), multiple disconnected regions with $\mathcal{F}=1$ appear, mirroring the fragmentation of the topological phase into several isolated islands that appear at $\mu=2.3$~meV.

All remaining horizontal cuts exhibit the same level of consistency between the Pfaffian and PDI predictions and the flux induced fermion parity switching captured by the FPSI.

\begin{figure}
    \centering
    \includegraphics[width=\linewidth]{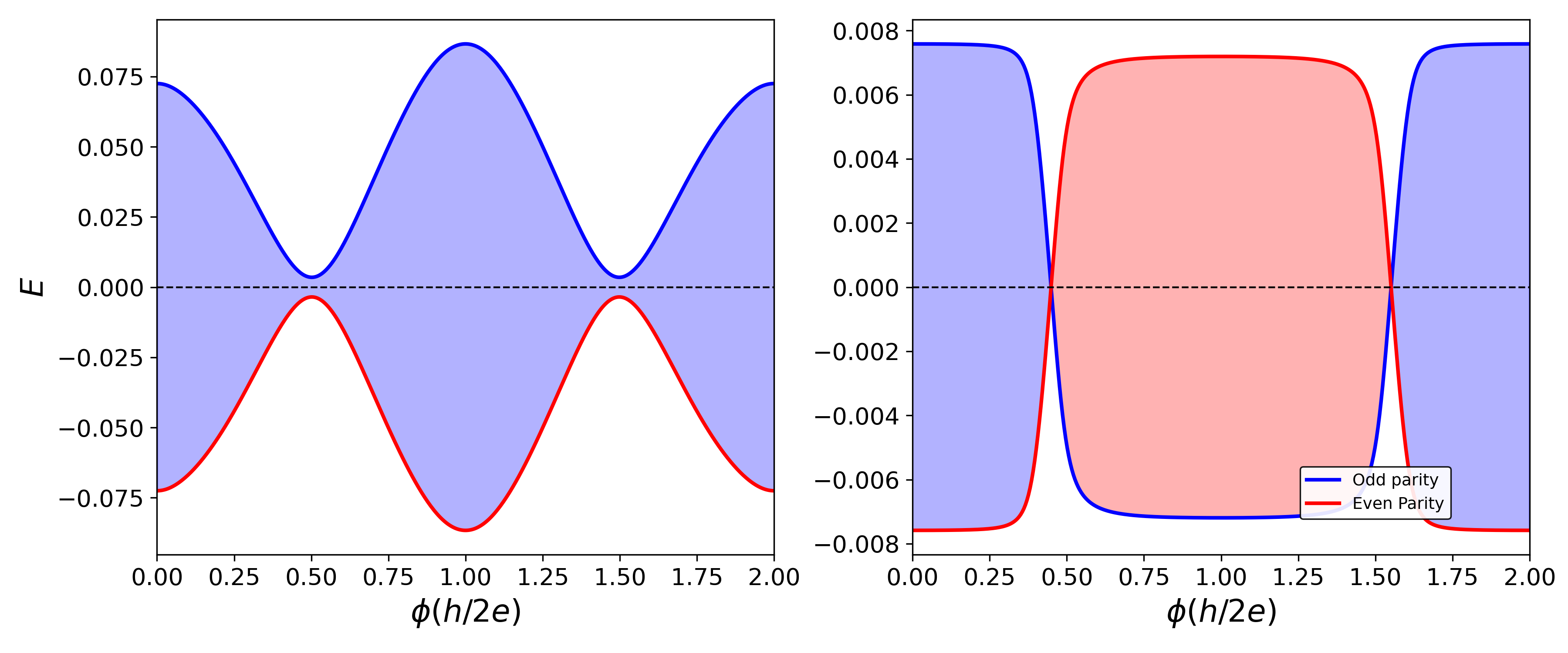}
    \caption{Flux dependent energy spectrum for the weakly disordered system, $V_0=0.5$~meV. The left panel shows the variation of 2 lowest BdG energy eigenvalues for a trivial region point ($\mu=2.3$~meV,$\Gamma=0.8$~meV) showing an avoided crossing and the right panel shows the behavior of the 2 energy curves for topological regime ($\mu=1.0$~meV,$\Gamma=0.8$~meV) indicating a clean singular crossing between $\Phi=0$ and $\Phi=h/2e$. The blue curve corresponds to odd parity while the red curve corresponds to even parity.}
    \label{fig:Evsphi_dis}
\end{figure}

Fig.~\ref{fig:Evsphi_dis} shows the flux dependent low energy spectrum for representative points taken from the weakly disordered system with disorder strength $V_0=0.5~\mathrm{meV}$. The evolution of the lowest BdG eigenvalues is plotted as a function of the flux induced phase $\Phi$. In the left panel, corresponding to a topologically trivial point at $(\mu=2.3~\mathrm{meV},\,\Gamma=0.8~\mathrm{meV})$, the spectrum exhibits an avoided crossing between the lowest energy states. As a consequence, the relative ordering of the even- and odd-parity energy branches remains unchanged between $\Phi=0$ and $\Phi=h/2e$, and no fermion parity switch occurs. By contrast, the right panel shows the spectrum for a point within the topological regime at $(\mu=1.0~\mathrm{meV},\,\Gamma=0.8~\mathrm{meV})$. In this case, a single, well defined zero energy crossing occurs between $\Phi=0$ and $\Phi=h/2e$. This leads to a sign change in both the even- and odd-parity energy branches over the interval, resulting in a switch of the ground state fermion parity from even to odd at $\Phi=h/2e$.

These flux induced spectral features demonstrate that both the real space twsited-boundary Pfaffian invariant and the periodic disorder invariant correctly capture fermion parity switching even in the presence of disorder.

\section{Conclusions}

In this work, we have presented a unified and physically transparent framework for Pfaffian-based topological invariants in one dimensional SM-SC hybrid systems. Beginning with Kitaev’s original momentum-space formulation in Sec.~\ref{sec:momentum_clean}, we reviewed how the $\mathbb{Z}_2$ invariant is determined entirely by the sign of the product of the Pfaffians of the Hamiltonian evaluated at particle–hole symmetric momenta. Applying this construction to the clean SM-SC nanowire model, we showed explicitly that the Pfaffian invariant reproduces the standard topological phase transition criterion $|\Gamma_c|=\sqrt{\mu^2+\Delta^2}$. Evaluated at $k=0$, the Pfaffian changes sign precisely at the bulk gap closing, thereby providing a direct algebraic diagnostic of the transition between trivial and topological phases. In this context, we also clarified why the determinant of the BdG Hamiltonian cannot serve as a topological invariant, as it eliminates the sign structure that encodes the relevant $\mathbb{Z}_2$ information.

We next reformulated Kitaev’s invariant in real space by establishing a correspondence between the Pfaffians of the Hamiltonian evaluated at particle–hole symmetric momenta and those computed under twisted boundary conditions in finite 1D SM-SC nanowire systems. Specifically, we considered a magnetic flux $\Phi$ threading a topological superconducting ring formed by joining the ends of the 1D SM-SC nanowire. Through the Peierls substitution, this flux manifests as a phase twist $\phi=\pi\Phi/\Phi_0$ in the hopping term connecting the two ends of the wire. As shown in Sec.~\ref{sec:relation_bw_ms_rs}, the product of Pfaffians evaluated under periodic ($\phi=0$) and anti-periodic ($\phi=\pi$) boundary conditions is equivalent to the product of Pfaffians at the particle–hole symmetric momenta. Within this formulation, the twisted boundary condition Pfaffian provides a natural real space representation of the $\mathbb{Z}_2$ invariant that remains well defined in finite systems.

To address spatial inhomogeneity and disorder, we introduced a superlattice construction in which the disordered segment is periodically repeated, thereby restoring discrete translational symmetry at the level of the superlattice (refer to Sec.~\ref{sec:PDI}). Within this framework, we formulated the Pfaffian invariant in terms of the superlattice Bloch momenta and showed that the sign of the product of the Pfaffians of the Hamiltonian evaluated at the particle–hole symmetric points $q=0$ and $q=\pi$ provides a $\mathbb{Z}_2$ topological invariant for spatially inhomogeneous systems. In the presence of chiral symmetry, we further demonstrated that this superlattice Pfaffian invariant is directly related to the periodic disorder invariant (PDI), with the parity of the PDI coinciding with the sign of the Pfaffian product at the particle–hole symmetric superlattice Bloch momenta.

Finally, by introducing a flux-induced phase twist in the superlattice boundary hopping, we showed that in the special case of a single supercell ($N_c=1$) the superlattice Bloch momentum reduces to the phase twist $q=\phi$, and the superlattice Pfaffian invariant becomes identical to the real-space twisted-boundary Pfaffian invariant defined through periodic and anti-periodic boundary conditions. This establishes that the twisted-boundary Pfaffian invariant for finite 1D SM-SC nanowire remains a well-defined $\mathbb{Z}_2$ topological invariant even in the presence of disorder. Complementing these structural results, we proved generally that the sign of the Pfaffian of a quadratic Hamiltonian coincides with the fermion parity of its ground state (refer to Sec.~\ref{sec:proof_fermion_parity}). The Pfaffian invariant and PDI may therefore be understood as a direct ground-state parity indicator.

Our numerical calculations illustrate that, in both clean and disordered finite systems, changes in the sign of the Pfaffian of the Hamiltonian evaluated at respective flux values and the PDI parity coincide with flux-induced level crossings and ground-state fermion parity switches (captured by FPSI $\mathcal{F}$ introduced in Sec.~\ref{sec:numerical_results}) in the low energy spectrum. 

Taken together, these results place the momentum-space Pfaffian invariant, its twisted-boundary-condition formulation, and the periodic-disorder invariant within a single coherent framework rooted in the fermion-parity of the ground state. This unified perspective clarifies the conceptual foundations of Pfaffian-based diagnostics and underscores their robustness in finite and spatially inhomogeneous nanowire systems. More broadly, the explicit connection between topological invariants, magnetic-flux response, and fermion-parity switching provides a concrete framework for spectroscopic and phase-sensitive measurements aimed at identifying topological superconductivity and Majorana physics in realistic devices.

\bigskip

\section{Acknowledgement}
We acknowledge support from ARO W911NF2210247, ONR-N000142312061, and SC-Quantum 2017750. We also thank J. D. Sau and T. D. Stanescu for fruitful discussions.

\bibliography{references.bib}

\end{document}